\documentclass[aps,pre,floats,twocolumn,showpacs,superscriptaddress]{revtex4}
\usepackage{graphicx}
\usepackage{amsmath}
\usepackage{mdwlist}

\newcommand{\comment}[1]{{}}

\newcommand{\beq}{\begin{equation}}
\newcommand{\eeq}{\end{equation}}
\newcommand{\ba}[1]{\begin{array}{#1}}
\newcommand{\ea}{\end{array}}
\newcommand{\bea}{\begin{eqnarray}}
\newcommand{\eea}{\end{eqnarray}}

\newcommand{\ben}{\begin{enumerate}}
\newcommand{\een}{\end{enumerate}}
\newcommand{\bit}{\begin{itemize}}
\newcommand{\eit}{\end{itemize}}
\newcommand{\bde}{\begin{description}}
\newcommand{\ede}{\end{description}}

\newcommand{\ds}{\displaystyle}

\newcommand{\sz}{\scriptsize}

\newcommand{\req}[1]{(\ref{#1})}

\newcommand{\avg}[1]{\langle {#1} \rangle}
\newcommand{\bdesc}[2]{\begin{basedescript}{\desclabelstyle{\pushlabel}\desclabelwidth{#1}\setlength{\labelsep}{0mm}\setlength{\leftmargin}{#2}}}
\newcommand{\edesc}{\end{basedescript}}

\begin{document}

\title{Non-perturbative heterogeneous mean-field approach to epidemic spreading in complex networks}

\author{Sergio G\'omez}
\affiliation{Departament d'Enginyeria Inform{\`a}tica i Matem{\`a}tiques,
  Universitat Rovira i Virgili,
  43007 Tarragona, Spain}

\author{Jes\'us G\'omez-Garde\~nes} \affiliation{Department of
  Condensed Matter Physics, University of Zaragoza, 50009 Zaragoza,
  Spain}

\affiliation{Institute for Biocomputation and Physics of Complex Systems (BIFI),
  University of Zaragoza, 50018 Zaragoza, Spain}

\author{Yamir Moreno}

\affiliation{Institute for Biocomputation and Physics of Complex
  Systems (BIFI), University of Zaragoza, 50018 Zaragoza, Spain}

\affiliation{Department of Theoretical Physics, Faculty of Sciences,
  University of Zaragoza, 50009, Zaragoza, Spain}

 \affiliation{Complex Networks and Systems Lagrange Lab, Institute for
   Scientific Interchange, Viale S. Severo 65, 10133 Torino, Italy}

 \author{Alex Arenas}

\affiliation{Departament d'Enginyeria Inform{\`a}tica i
  Matem{\`a}tiques, Universitat Rovira i Virgili, 43007 Tarragona,
  Spain}

\affiliation{Institute for Biocomputation and Physics of Complex
  Systems (BIFI), University of Zaragoza, 50018 Zaragoza, Spain}

\date{\today}

\begin{abstract}
Since roughly a decade ago, network science has focussed among others
on the problem of how the spreading of diseases depends on structural
patterns. Here, we contribute to further advance our understanding of
epidemic spreading processes by proposing a non-perturbative
formulation of the heterogeneous mean field approach that has been
commonly used in the physics literature to deal with this kind of
spreading phenomena. The non-perturbative equations we propose have no
assumption about the proximity of the system to the epidemic
threshold, nor any linear approximation of the dynamics.  In
particular, we first develop a probabilistic description at the node
level of the epidemic propagation for the so-called
susceptible-infected-susceptible family of models, and after we derive
the corresponding heterogeneous mean-field approach. We propose to use
the full extension of the approach instead of pruning the expansion to
first order, which leads to a non-perturbative formulation that can be
solved by fixed point iteration, and used with reliability far away
from the epidemic threshold to assess the prevalence of the epidemics.
Our results are in close agreement with Monte Carlo simulations thus
enhancing the predictive power of the classical heterogeneous mean
field approach, while providing a more effective framework in terms of
computational time.
\end{abstract}


\maketitle

\section{Introduction}

During the last decade, the physicists' community working on the
theory of complex networks has paid special attention to the problem
of epidemic spreading in social \cite{n02}, biological
\cite{hethcote,Maybook,daley,Murray} and technological networks
\cite{vespbook08}. The development of mathematical and computational
models to guide our understanding of the disease dynamics has allowed
to address important issues such as the influence of diverse contact
patterns and also new algorithms for immunization and vaccination
policies \cite{Maybook,geisel,egamstw04}. Physicist's approaches to
problems in epidemiology invoke statistical physics, the theory of
phase transitions and critical phenomena \cite{stanley}, to grasp the
macroscopic behavior of epidemic outbreaks
\cite{vespiromu,vrpre,llm01,mpsv02,n02,bbpsv04,glmp08,cwwlf08}. It is
not adventurous to claim that one of the main artifices of this
success has been the Mean-Field (MF) approximation, where homogeneity
and isotropy are hypothesized to reduce the complexity of the system
under study.

On the other hand, the study of the topological properties of complex
networks \cite{newmanrev,yamirrep,dgm08} has provided new grounds to
the understanding of contagion dynamics. Particularly widespread in
nature are heavy tailed degree distributions, specially scale-free
(SF) networks, whose degree distribution follows a power law $P(k)\sim
k^{-\gamma}$ for the number of connections, $k$, an individual has. SF
networks include patterns of sexual contacts \cite{stanley_nat}, the
Internet \cite{vespbook08}, as well as many other social,
technological and biological networks \cite{caldarelli}. The critical
properties of an epidemic outbreak in SF networks were addressed using
the heterogeneous mean-field (HMF) prescription (also called
correlated mean field)
\cite{vespiromu,vrpre,llm01,mpsv02,n02,bbpsv04,glmp08}. HMF
coarse-grains vertices within degree classes and hypothesizes that all
nodes in a degree class have the same dynamical properties; the
approach also assumes that fluctuations can be neglected
\cite{vespiromu}.  This framework has been proved to be exact in
annealed networks, whose nodes' degrees are sampled from a fixed
degree distribution at each step of the dynamics \cite{dgm08}
(i.e.\ its specific connectivity is fixed only in average). However,
in specific realizations of a model's topology (quenched networks),
HMF can result in different levels of accuracy \cite{Guerra}. This
problem leads to the question of whether or not the direct use of the
HMF approach is accurate enough when dealing with real networks (i.e.,
in an instance of a network ensemble). In a recent work, Gleeson et
al.\ \cite{gleeson} studied how accurate the HMF can be on 21
real-world networks and found some relationships between the
predictive accuracy and topological properties of the networks. Also
other works have addressed this problem using pair approximations, in
particular Miller \cite{mill} has found analytical results for a
special class of clustered networks.

Although the HMF approach has been extremely useful to assess the
critical properties of epidemic models, it is not thought to give
information of individual nodes but of classes of nodes of a given
degree. Then, asking about the probability that a given node is
infected is not well-posed in this framework. To obtain more details
at the individual (node) level of description, usually one has to rely
on Monte Carlo (MC) simulations of the actual dynamics, which have
also been used to validate the results obtained using HMF methods. The
current theory concentrates in two specific situations, the contact
process
\cite{Marro,Castellano:2006,HHP,Castellano:2007,Castellano:2008,marian}
(CP) and the fully reactive process \cite{GA04,cbp05,cpv07} (RP). A CP
stands for a dynamical process that involves an individual stochastic
contagion per infected node per unit time, while in the RP there are
as many stochastic contagions per unit time as neighbors a node
has. However, in real situations, the number of stochastic contacts
per unit time is surely a variable of the problem itself
\cite{meloni}.

Recently, some of us have proposed an alternative approach to study
the Susceptible-Infected-Susceptible (SIS) model
\cite{hethcote,daley,Murray} considering the number of stochastic
contacts as a parameter that defines a whole family of SIS models
\cite{EPLnostre}, the so-called Microscopic Markov-Chain Approach
(MMCA). This more realistic scenario allows to characterize the
prevalence of the disease at the level of individual nodes when the
number of contacts interpolates between the two limiting cases of CP
and RP.  Capitalizing on the MMCA framework, we propose here a
non-perturbative HMF formulation which does not prune the equations at
first order in the prevalence density, but considers the whole series
expansion. The resulting equations can be solved using fixed point
iteration, and are accurate in predicting the epidemic incidence for
the whole phase diagram, even out of the critical transition region.

\section{Discrete-time and Continuous-time SIS models in networks \label{sect2}}

The analysis of epidemic spreading models in networks has some of the same difficulties that are found for well-mixed populations (equivalent to complete graphs). These difficulties are intrinsic to the discrete-time or continuous-time formulation of the governing equations, and the methods used to solve each of them. Continuous approximations have been more popular in epidemic modeling because of their mathematical tractability, and the avoidance of chaotic behaviors that can arise in their discrete counterparts \cite{allen}. For the sake of clarity, we henceforth fix our attention on the study of a family of SIS models. In a SIS model, individuals that are cured do not develop permanent immunity but are immediately susceptible to the disease again. In well-mixed populations, the differential equations governing the number of susceptible (S) and infected (I) individuals are
\begin{eqnarray}
  \frac{dS}{dt} &=&\ds  - \tilde{\beta} S \frac{I}{N} + \tilde{\mu} I\,, \nonumber \\
  \frac{dI}{dt} &=&\ds  \tilde{\beta} S \frac{I}{N} - \tilde{\mu} I\,,
  \label{cont}
\end{eqnarray}
where $N = S(t) + I(t)$ is the (constant) size of the population. The term $I/N$ accounts for the probability of contacting an infected individual in a well-mixed population of size $N$, $\tilde{\beta}$ is the infectivity rate (probability per unit time) for each contact, and $\tilde{\mu}$ is the rate at which one infected individual recovers. Their corresponding difference equations are
\begin{eqnarray}
  S(t+\Delta t) &=&\ds S(t) \ \Big( 1- \tilde{\beta} \Delta t \frac{I(t)}{N} \Big) + \tilde{\mu} \Delta t I(t)\,, \nonumber \\
  I(t+\Delta t) &=&\ds I(t) \Big[ 1- \tilde{\mu} \Delta t + \tilde{\beta} \Delta t\frac{I(t)}{N} S(t) \Big]\,,
  \label{diff}
\end{eqnarray}
or equivalently
\begin{equation}
  I(t+\Delta t) = I(t) - \tilde{\mu} \Delta t I(t) + \tilde{\beta} \Delta t \frac{I(t)}{N} [N - I(t)]\,.
  \label{diffi}
\end{equation}
Defining $\rho(t)=I(t)/N$ as the fraction of infected individuals in the population, Eq.~(\ref{diffi}) is written as
\begin{equation}
  \rho(t+\Delta t) = \rho(t) - \tilde{\mu} \Delta t\, \rho(t) + \tilde{\beta} \Delta t\, \rho(t) [1 - \rho(t)]\,.
  \label{difffraci}
\end{equation}

Note that while the system of Eqs.~(\ref{cont}) always converges to a
solution \cite{heth76,allen}, Eq.~(\ref{difffraci}) can be mapped to a
logistic function when the reproductive ratio ${\cal R} =
\tilde{\beta}/\tilde{\mu}>1$, thus giving rise to basic periodicity,
bifurcations and chaotic behavior depending on the parameters
\cite{allen}. Although both descriptions are equivalent in the limit
$\Delta t \rightarrow 0$, differences arise when considering a finite
$\Delta t$.

Particularly interesting is what happens when considering a numerical
scheme iterating Eq.~(\ref{difffraci}). In many cases $\Delta t$ is
usually assimilated to the stochastic simulation time unit and set to
1. Consequently, the numerical differences with the continuous case
are substantial. It is also important to distinguish between rates and
probabilities, $\tilde{\beta} \Delta t= \beta$ is a probability, and
the same holds for $\tilde{\mu} \Delta t = \mu$. Again, by setting
$\Delta t=1$, one can mix up rate and probabilities because both will
have the same values, even though their units are different.

The mapping of the above SIS equations to the case of heterogeneous networks is not straightforward and has its critical step in the redefinition of the probability of contacts. In a network, the number of contacts is restricted to a fixed neighborhood, then each individual (node) can potentially contact only its neighbors. In the seminal work by Pastor-Satorras and Vespignani \cite{vespiromu}, these authors proposed the direct use of Eq.~(\ref{difffraci}) for classes of nodes based on their degree $k$. This is the root of the HMF approach in complex networks and the hypothesis of homogeneity is here postulated at the level of classes of nodes. The rationale behind this assumption is that the dynamical behavior of any two nodes with the same degree $k$ will be essentially the same. Then, a system of equations for each class $k$ is written as
\begin{equation}
  \rho_{k} (t+\Delta t)= \rho_{k} (t) - \tilde{\mu} \Delta t\, \rho_{k} (t) + \tilde{\beta} \Delta t\, \Theta_{k} (t) [1 - \rho_{k} (t)]\,,
  \label{HMFvr}
\end{equation}
where now $ \rho_{k} (t)$ stands for the fraction of infected individuals of degree $k$, and the probability of contacting an infected node is encoded in the new function $\Theta_{k} (t)$. For the general case of correlated networks, the function $\Theta_{k} (t)$ takes the form
\begin{equation}
  \Theta_{k}(t) = \sum_{k'} P(k'|k) \rho_{k'}\,,
  \label{teta}
\end{equation}
where $P(k'|k)$ is the probability that a node of degree $k$ connects to a node of degree $k'$. Eq.(\ref{HMFvr}) is used to find the stationary value of the incidence for given values of $\beta$ and $\mu$. Indeed, for the stationary state, it is true that the only dependence to take into account is the ratio $\lambda=\beta/\mu$ because the equations can be rescaled without modifying their solutions. However, this is not anymore the case during the transient. The critical value $\lambda_c$ for the epidemic threshold was found in \cite{bogupas} to be
\begin{equation}
  \lambda_c=\frac{1}{\Lambda_{\mbox{\sz max}}(C)}\,,
  \label{lambc}
\end{equation}
being $\Lambda_{\mbox{\sz max}}(C)$ the largest eigenvalue of the connectivity matrix of classes of nodes $C$, whose components are given by $C_{kk'}=kP(k'|k)$, i.e.\ the expected number of links from a node of degree $k$ to nodes of degree $k'$.

\section{The Microscopic Markov-Chain Approach to SIS models}

In this section, we summarize an alternative approach to describe the equations governing the SIS class of models \cite{EPLnostre}. We henceforth refer to this formulation as Microscopic Markov-Chain Approach (MMCA). The main advantage of this approach is that it is able to deal with the infection dynamics at the level of single nodes. Specifically, we focus on the probability of a node to be infected at time $t$. For each node $i$, we construct a Markov chain that accounts for the probability of infection $p_i(t)$, assuming that the number of contacts of node $i$ with its neighbors is parameterized by an integer $\eta$, and that the infection events are uncorrelated. Note that this constitutes a first principles derivation of a discrete model, not a discretization of a differential equation. We consider two different situations of interest: i) without reinfections (WOR), and ii) with reinfections (WIR). The first case implies that the time scales for the infection and cure are well separated, whereas the latter assumes that the same time scale holds for infection and cure and therefore a just recovered individual might catch the disease again within the same time step $t$. The respective equations are:
\begin{alignat}{2}
  &\mbox{WOR:} \nonumber
  \\
  &p_{i}(t+1) = [1-p_{i}(t)][1-q_{i}(t)] + (1-\mu) p_{i}(t)\,,
  \label{eqwor}
  \\ \nonumber
  \\
  &\mbox{WIR:}& \nonumber
  \\
  &p_{i}(t+1) = [1-q_{i}(t)] + (1-\mu) p_{i}(t) q_{i}(t)\,,
  \label{eqwir}
\end{alignat}
where the probability $q_i(t)$ of node $i$ not being infected by any neighbor is
\begin{equation}
  q_i(t) = \prod_{j=1}^{N} [1-\beta r_{ji} p_j(t)]\,.
  \label{eqqi}
\end{equation}

Keeping in mind the separation of the two processes, namely,
contacting a node and transmitting the infection, already presented in
Eqs.~(\ref{diff}) for the well-mixed case, the explanation of these
equations is straightforward. The terms in the r.h.s.\ of
Eq.~(\ref{eqwor}) account respectively for the probability that a
susceptible node $[1 - p_i(t)]$ is infected by at least one neighbor
$[1 - q_i(t)]$, and an infected node does not recover $[(1 -
  \mu)p_i(t)]$. Eq.~(\ref{eqwir}) adds, after some algebra, a term
that accounts for the probability that an infected node recovers $[\mu
  p_i(t)]$ but gets infected again by a neighbor $[1 - q_i(t)]$ in the
same time step. Finally, in Eq.~(\ref{eqqi}), we have the
probabilities that infected nodes $[p_j(t)]$ contact node $i$, and
that these contacts lead to new infections, which occur with
probability $\beta$. The values of the contact probabilities $r_{ji}$
can be expressed as
\begin{equation}
  r_{ji} = R_{\eta}\left(\frac{w_{ji}}{w_j}\right)
  \label{rjiw}
\end{equation}
for weighted networks, and
\begin{equation}
  r_{ji} = R_{\eta}\left(\frac{a_{ji}}{k_j}\right) = a_{ji} R_{\eta}\left(\frac{1}{k_j}\right)
  = a_{ji} R_{\eta}(k_j^{-1})
  \label{rjiu}
\end{equation}
for unweighted networks, where
\begin{equation}
  R_{\eta}(x) = 1 - (1 - x)^{\eta}\,.
\end{equation}
For the contact process, $\eta=1$ and $R_{1}(x)=x$, whereas for the fully reactive process,
$\eta\rightarrow\infty$ and $R_{\infty}(x)=1$

At the stationary state, Eqs.~(\ref{eqwor}) and~(\ref{eqwir}) are independent of the discrete time-step, and simplify to
\begin{alignat}{2}
  &\mbox{WOR:}& \ \ &p_{i} = (1-p_{i})(1-q_{i}) + (1-\mu) p_{i}\,,
  \label{eqworst}
  \\
  &\mbox{WIR:}& \ \ &p_{i} = (1-q_{i}) + (1-\mu) p_{i} q_{i}\,,
  \label{eqwirst}
\end{alignat}
with
\begin{equation}
  q_i = \prod_{j=1}^{N} (1-\beta r_{ji} p_j)\,.
  \label{eqqst}
\end{equation}
These equations are easily solved by fixed point iteration until a fixed point (for the WIR and WOR cases) or a cycle (for the WOR setup) is found. In this second case, averaging over the oscillating values of the quantity of interest must be considered. Finally, the average fraction of infected nodes in the stationary state is given by
\begin{equation}
  \rho = \frac{1}{N}\sum_{i=1}^{N} p_i\,.
\end{equation}

\section{Non-perturbative HMF}

In heterogeneous mean field theory it is supposed that all nodes of the same degree behave equally.
In terms of the MMCA formulation this means that $p_i = p_j$ if $k_i=k_j$, and the
density $\rho_k$ of infected nodes of degree $k$ is given by
\begin{equation}
  \rho_k = \frac{1}{N_k} \sum_{j\in K} p_j = p_i\,,\ \ \forall i\in K\,,
  \label{rhok}
\end{equation}
where $K$ is the set of nodes with degree $k$, whose cardinality is denoted by $N_k$.
This notation allows to group together terms according to the degrees of the nodes.
For instance, if the degree of node $i$ is $k_i=k$, then
\begin{align}
  \sum_j a_{ji} p_j
    &= \sum_{k'} \sum_{j\in K'} a_{ji} \rho_{k'}
    = \sum_{k'} \rho_{k'} \sum_{j\in K'} a_{ij}
  \nonumber \\
    &= \sum_{k'} \rho_{k'} C_{kk'}
    = k \sum_{k'} P(k'|k) \rho_{k'}\,,
  \label{exsum}
\end{align}
where $C_{kk'} = k P(k'|k)$ is the expected number of links from a node of degree $k$ to nodes of degree $k'$, as explained in Sect.~\ref{sect2}.

Substitution of the HMF approximation, Eq.~(\ref{rhok}), into the MMCA
Eqs.~\req{eqworst} and~\req{eqwirst} leads to
\begin{alignat}{2}
  &\mbox{WOR:}& \ \ \rho_{k} &= (1-\rho_{k})(1-q_{k}) + (1-\mu) \rho_{k}\,,
  \label{nphmfworiter}
  \\
  &\mbox{WIR:}& \ \ \rho_{k} &= (1-q_{k}) + (1-\mu) \rho_{k} q_{k}\,,
  \label{nphmfwiriter}
\end{alignat}
which can also be written as
\begin{alignat}{2}
  &\mbox{WOR:}& \ \ 0 &= -\mu \rho_k + (1-\rho_{k})(1-q_{k})\,,
  \label{nphmfwor}
  \\
  &\mbox{WIR:}& \ \ 0 &= -\mu \rho_k + (1-(1-\mu)\rho_{k})(1-q_{k})\,.
  \label{nphmfwir}
\end{alignat}
These equations constitute the HMF approximations of MMCA for the SIS model without and with reinfections, respectively.

We still need a HMF expression for $q_k$. For unweighted networks the value of $q_i$ is
\begin{equation}
  q_i = \prod_{j=1}^{N} (1-\beta r_{ji} p_j) = \prod_{j=1}^{N} \left(1-\beta a_{ji} R_{\eta}(k_j^{-1}) p_j\right)\,.
\end{equation}
If node $i$ has degree $k_i=k$, then
\begin{equation}
  q_i = q_k = \prod_{k'}\prod_{j\in K'} \left(1-\beta a_{ji} R_{\eta}(k'^{-1}) \rho_{k'}\right)\,,
\end{equation}
where we have grouped together the terms in the product by their degrees $k'$ as in Eq.~(\ref{exsum}). The expression within the parentheses is equal to $1$ for all nodes of degree $k'$ that are not connected to node $i$ ($a_{ji}=0$), and equal to $[1-\beta R_{\eta}(k'^{-1}) \rho_{k'}]$ for nodes of degree $k'$ that are linked to $i$ ($a_{ji}=1$).  Besides, the expected number of such terms is $C_{kk'}$. Hence, we obtain
\begin{equation}
  q_k = \prod_{k'} \left(1-\beta R_{\eta}(k'^{-1})
  \rho_{k'}\right)^{C_{kk'}}\,.
  \label{nphmfq}
\end{equation}
Eqs.~\req{nphmfwor} and~\req{nphmfwir}, together with
Eq.~\req{nphmfq}, form what we call the {\em Non-perturbative
  Heterogeneous Mean Field} (npHMF) equations of the SIS model in
unweighted networks. Note that in the derivation of the npHMF
equations no assumption has been made about the proximity of the
system to the epidemic threshold, where the epidemic prevalence is
small, nor any linear approximation has been invoked, hence the
qualification of ``non-perturbative''.

The solution of the npHMF equations follows the same steps as in MMCA,
i.e., Eqs.~(\ref{nphmfworiter}), (\ref{nphmfwiriter})
and~(\ref{nphmfq}) are iterated until a fixed point (WIR, WOR) or a
cycle (WOR) is found. As before, for the latter case, we take the
average of oscillating values for the disease prevalence. Finally, the
global epidemic prevalence is given by
\begin{equation}
  \rho = \frac{1}{N}\sum_{k} N_k \rho_k\,.
\end{equation}

It is easy to show that the standard HMF equations \cite{bogupas,marian} are just a linear approximation of our WOR npHMF equations. Near the epidemic threshold $\beta_c$, where $\rho_k \ll 1$, we get
\begin{align}
  q_{k} &\sim  1-\beta \sum_{k'} C_{kk'} R_{\eta}(k'^{-1}) \rho_{k'}
  \nonumber\\
  &=  1-\beta k \sum_{k'} P(k'|k) R_{\eta}(k'^{-1}) \rho_{k'}\,,
\end{align}
which can be inserted into Eq.~\req{nphmfwor} to give
\begin{equation}
  0 = -\mu \rho_k + \beta k (1-\rho_{k}) \sum_{k'} P(k'|k) R_{\eta}(k'^{-1}) \rho_{k'}\,,
\end{equation}
where
\begin{equation}
  R_{\eta}(k'^{-1}) =
  \begin{cases}
     1  & \mbox{for the RP,} \\
     \ds \frac{1}{k'}  & \mbox{for the CP.} \\
  \end{cases}
\end{equation}

\subsection{Epidemic threshold}

To round off our analysis, we derive the critical spreading threshold using the different approaches here discussed. In~\cite{EPLnostre} it is shown that MMCA allows the determination of the epidemic threshold
\begin{equation}
  \beta_c = \frac{\mu}{\Lambda_{\mbox{\sz max}}(R)}\,,
\end{equation}
where $\Lambda_{\mbox{\sz max}}(R)$ is the maximum eigenvalue of the matrix $R$ of the contact probabilities $r_{ij}$. In particular,
\begin{equation}
  \beta_c =
  \begin{cases}
     \ds \frac{\mu}{\Lambda_{\mbox{\sz max}}(A)}  & \mbox{for the RP,} \\
     \mu  & \mbox{for the CP.} \\
  \end{cases}
\end{equation}
These results hold both with and without reinfections, since at first order, Eqs.~\req{eqwor} and~\req{eqwir} coincide. In the same way, the critical points from npHMF Eqs.~\req{nphmfwor} and~\req{nphmfwir} are the same as in standard HMF, where the matrix $H$ with elements $h_{kk'}=C_{kk'}R_{\eta}(k'^{-1})$ replaces matrix $R$,
\begin{equation}
  \beta_c^{\mbox{\sz HMF}} = \frac{\mu}{\Lambda_{\mbox{\sz max}}(H)}\,,
\end{equation}
with the well-known particular cases
\begin{equation}
  \beta_c^{\mbox{\sz HMF}} =
  \begin{cases}
     \ds \frac{\mu}{\Lambda_{\mbox{\sz max}}(C)}  & \mbox{for the RP,} \\
     \mu  & \mbox{for the CP.} \\
  \end{cases}
\end{equation}
The comparison with Monte Carlo (MC) simulations shows that MMCA provides a better approximation to the epidemic threshold than HMF, since no information of the original network structure is lost. Additionally, by the definition of the contact probabilities, the MMCA is applicable to weighted and directed networks without further modifications of the equations. This is not the case when one deals with HMF in networks that are not unweighted and undirected.

\begin{figure}[!t]
  \begin{center}
  \includegraphics*[width=0.49\textwidth,angle=0]{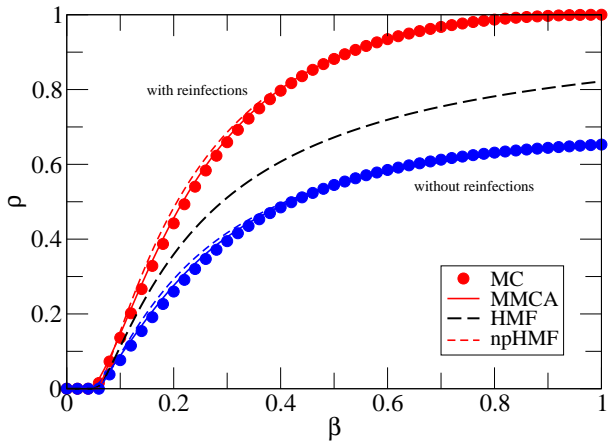}
  \end{center}
  \caption{Epidemic prevalence $\rho$ as a function of the infection rate $\beta$. The different curves correspond (as indicated) to Monte Carlo results and the numerical solutions of the fixed point equations of the different approaches (MMCA, HMF and npHMF) for a SF network made up of $N=10^4$ nodes using the configuration model. The exponent of the degree distribution is $\gamma=2.7$. The contact probabilities used correspond to the case of a fully reactive process and $\mu$ has been set to $0.5$.}
  \label{fig1}
\end{figure}

\section{Comparison with Monte Carlo simulations}

To compare the results coming out from the different approaches, we
have performed MC simulations of the disease dynamics on top of
scale-free networks generated using the uncorrelated configuration
model. Besides, numerical solutions of the fixed point equations have
also been obtained. For high values of $\mu$ and $\beta$, MMCA without
reinfections does not converge to a fixed point as the rest, but it
converges to an oscillation between two states, from which the average
is taken. These oscillations are also present in the MC runs, and
dissapear after averaging over multiple runs. On the other hand,
standard HMF cannot be solved by iteration, since it diverges even for
small values of $\beta\sim 10^{-1}$. Therefore, we have made use of a
nonlinear minimization algorithm based on a successive quadratic
programming solver.

\begin{figure}[tbc]
  \begin{center}
  \includegraphics*[width=0.49\textwidth,angle=0]{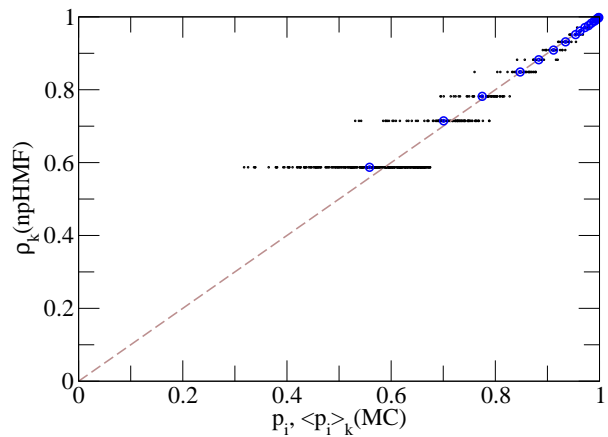}
  \end{center}
  \caption{Fraction $\rho_k$ of infected individuals of degree $k$ from npHMF versus the infection probabilities $p_i$ (black) and their average $\avg{p_i}_k$ over nodes with the same degree (blue) obtained from MC. The epidemic model is a RP with reinfections with $\mu=0.5$ and $\beta=0.3$. For the sake of clarity, we have used in this case an SF network of 500 nodes with $\gamma=2.7$.}
  \label{fig2}
\end{figure}

Figure\ \ref{fig1} shows the comparison between all the approaches
discussed throughout this paper for the case in which a fully reactive
process is considered. As can be seen, the worst performance with
respect to MC simulations corresponds to the HMF, which correctly
predicts the epidemic threshold but fails to reproduce the evolution
of the epidemic incidence as the infection rate increases and moves
away from the critical point. The figure also shows that the npHMF
approximation behaves only slightly worse than the Markov Chain
formulation. More important, in addition to correctly capturing the
critical epidemic threshold and at variance with the standard HMF, it
allows to study the whole phase diagram whatever the value of the
infection rate is.

\begin{figure}[tbc]
  \begin{center}
  \begin{tabular}{c}
  \mbox{\includegraphics*[width=0.49\textwidth,angle=0]{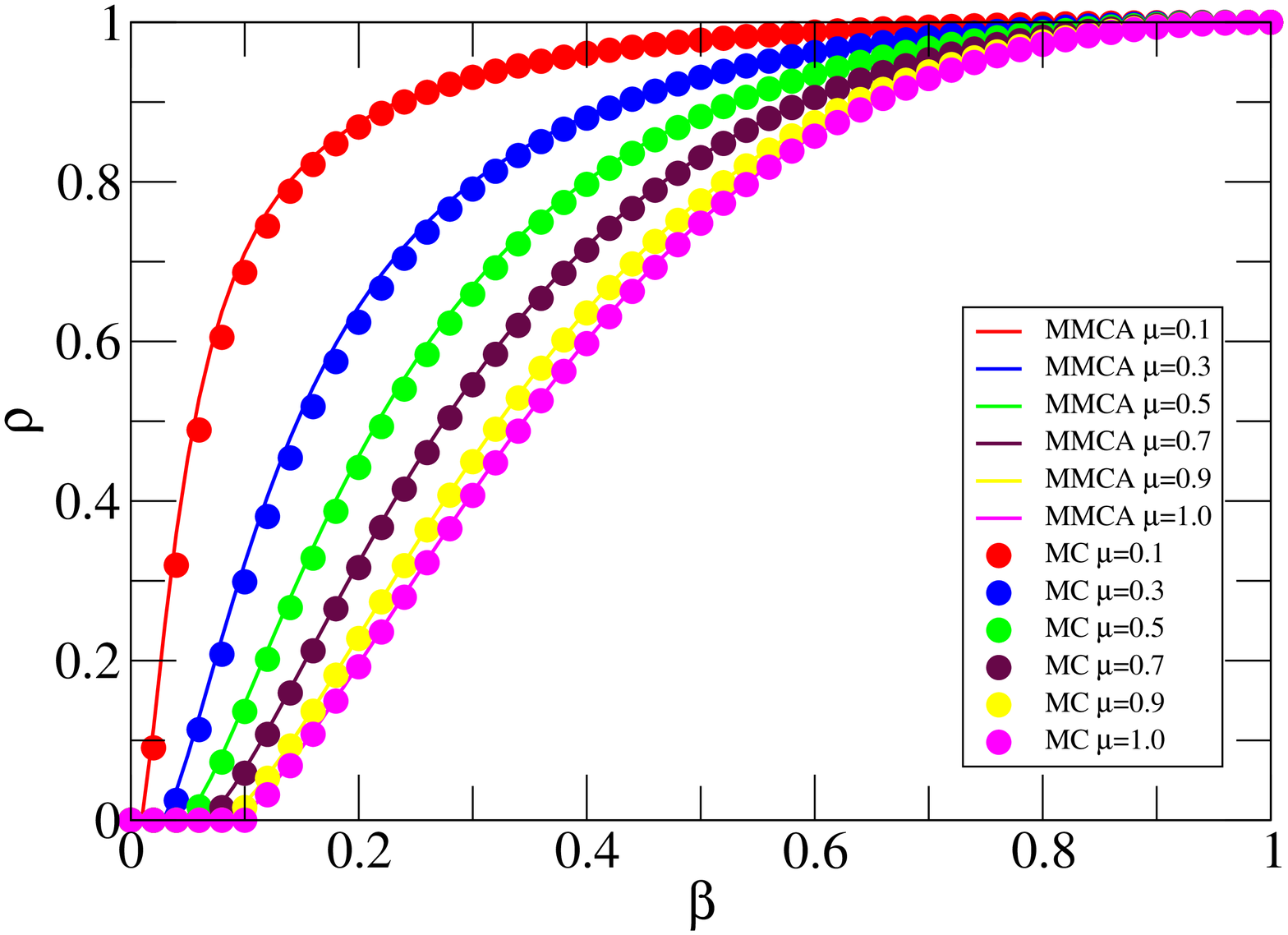}}
  \\
  \mbox{\includegraphics*[width=0.49\textwidth,angle=0]{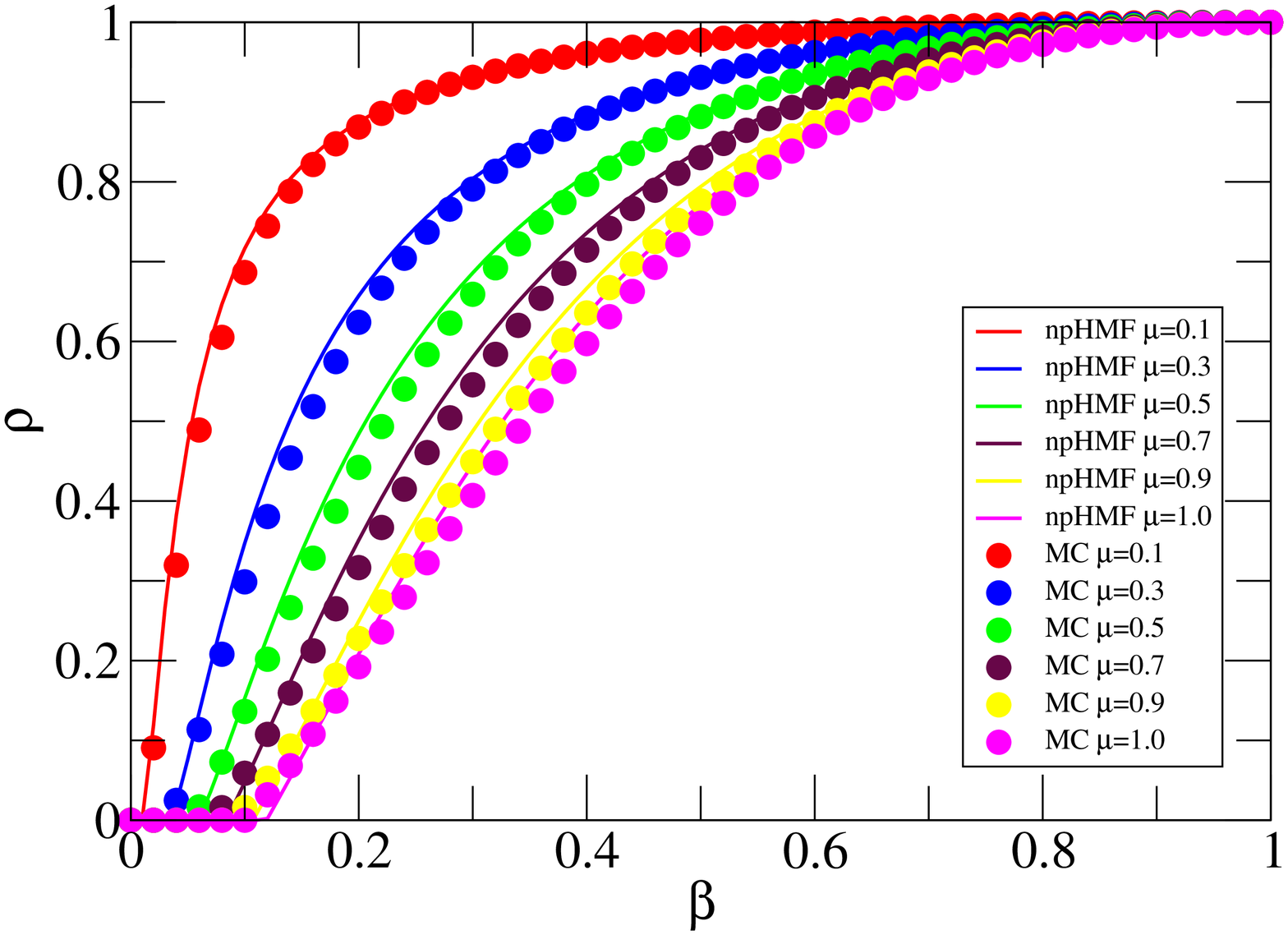}}
  \end{tabular}
  \end{center}
  \caption{Epidemic prevalence $\rho$ as a function of the infection rate $\beta$ for different values of $\mu$ as indicated. The top panel corresponds to the numerical solutions of the fixed point equations of the MMCA formulation, while the bottom figure has been obtained using the npHMF equations. In both cases, the numerical solutions are compared with the results of MC simulations on top of a SF network made up of $N=10^4$ nodes. The exponent of the degree distribution is $\gamma=2.7$. The contact probabilities used correspond to the case of a fully reactive process.}
  \label{fig3}
\end{figure}

To provide further evidences of the validity of the npHMF, we have
represented in Fig\ \ref{fig2} the epidemic incidence of nodes of
degree $\rho_k$ as given by the solution of the npHMF equations as a
function of the probability $p_i$ that a node $i$, whose connectivity
is $k$, is infected, being the latter obtained from MC
simulations. Besides, the average over all the $p_i$'s for nodes of
the same degree has also been represented. As can be seen, all
individual probabilities are distributed around the mean value, which
in turns is close to the values coming out from the npHMF
formulation. These results further illustrate that the behavior is not
the same along all the connectivity classes: for large prevalence
densities (or probability of being infected), the dispersion around
the mean values shrinks. In other words, the degree of accuracy in the
prediction of the state of a given node $i$ with respect to whether it
is infected or not depends on $\rho_k$: for large values of this
density, one can predict with high accuracy the individual
probabilities $p_i$ for nodes of degree $k$, while if $\rho_k$ is
relatively small, the prediction error is larger due to the more
pronounced deviations from the average value.

Finally, we have also tested the new formulation against variation of
the recovery rate. The results obtained for different values of $\mu$
using the MMCA and the npHMF approaches are compared with MC
simulations in Fig.\ \ref{fig3}. As in Fig.\ \ref{fig1}, we have
prescribed a fully reactive process on top of an undirected and
unweighted scale-free network. Moreover, the results shown correspond
to the case in which reinfections are possible (similar results are
found for the case without reinfections). The different curves show
that both formulations perform quite well, being however the MMCA
slightly better that the npHMF case.

\section{Conclusions}

In summary, in this paper we have proposed a new kind of heterogeneous
mean field approach to describe the spreading of diseases in complex
networks. Capitalizing on a previous formulation aimed at computing
the individual probabilities for nodes to be infected, we have coarse
grained the dynamical equations by degree classes. In doing so, we
have been able to keep higher order terms, which ultimately allow us
to use the resulting fixed point equation to obtain numerical
solutions that are in good agreement with MC simulations. However,
although the results outperform those obtained through the standard
mean-field approach, the new formulation is still limited with respect
to individual-based approaches (such as the MMCA). The reason is that,
when coarse-graining the dynamics, some information of the original
network is lost and exact quantities are replaced by expected
values. The ultimate consequence is that, although one can accurately
reproduce the behavior of global dynamical descriptors like the
epidemic prevalence, at the individual level the probabilities of
infection are more error-prone. Having said that, we however think
that the approach proposed here represents a significant improvement
on the standard HMF framework in several aspects. First, it is more
accurate. Secondly, it allows to solve the whole phase diagram, which
opens up the possibility of getting fast numerical solutions given a
network topology without resorting to MC simulations.


\begin{acknowledgments}
We acknowledge Prof. Mason A. Porter for the critical reading of our
manuscript.  Partial support came from the Spanish MICINN projects
FIS2009-13730-C02-02 (S.G. and A.A.), FIS2008-01240 (J.G.G. and Y.M.),
MTM2009-13848 (J.G.G.) and FIS2009-13364-C02-01 (Y.M.). S. G. and
A. A. were partially supported by the Generalitat de Catalunya
2009-SGR-838. Y. M. was partially supported by the FET-Open project
DYNANETS (grant no. 233847) funded by the European Commission and by
Comunidad de Arag\'on (Spain) through the project
FMI22/10. J.~G.-G.~acknowledges support from the MICINN (Spain)
through the Ram\'on y Cajal Program.
\end{acknowledgments}

\end{document}